\documentclass[aps, twocolumn,floatfix,superscriptaddress,longbibliography,prx]{revtex4-2}
\usepackage[utf8]{inputenc}
\usepackage{natbib}
\usepackage[normalem]{ulem}
\usepackage{graphicx}
\usepackage{xcolor}
\UseRawInputEncoding

\usepackage[tbtags]{amsmath}
\usepackage{hyperref}
\hypersetup{colorlinks,allcolors=black}
\usepackage{amssymb}
\usepackage{textcomp,gensymb}
\usepackage{float}
\usepackage{amsmath}
\usepackage{tabularx,graphicx}
\usepackage{epstopdf}
\usepackage{latexsym}
\usepackage{color, colortbl}
\usepackage{psfrag}
\usepackage{bbm}
\usepackage{bm}
\usepackage{titlesec}
\usepackage{dsfont}
\usepackage{feynmp}
\usepackage{slashed}
\usepackage{multirow}
\usepackage{soul} 
\usepackage{braket}
\usepackage[margin=10pt,font=small,labelfont=bf,
labelsep=endash,justification=justified]{caption}
\usepackage{ragged2e}  
\DeclareCaptionJustification{justified}{\justifying}
\usepackage{pgfplots}
\usepackage{siunitx}
\usepackage[Export]{adjustbox} 
\adjustboxset{max size={0.9\linewidth}{0.9\paperheight}}
 \usepackage{commath}
 \usepackage{mathtools}
\usepackage{listings}
 \usepackage{comment}
 \usepackage{wrapfig}
 \usepackage{subfiles}
 \usepackage{multirow}
\usetikzlibrary{plotmarks}

\DeclareMathOperator{\diag}{Diag}

\renewcommand{\hat}[1]{{\bf {\widehat #1}}}
\renewcommand{\phi}{\varphi}
\renewcommand{\epsilon}{\varepsilon}

\def\H#1{\mathrm{H}_{\text{#1}}}

\def\Auc{A_{\rm{uc}}}
\def\hsp{H_{\text{kin}}}

\def\intstrengthfock{U_{\text{Fock}}}

\hypersetup{colorlinks=true,linkcolor=blue,citecolor=blue,urlcolor=blue}

\pgfplotsset{compat=1.18}
\begin{document}

\title{Single-gate tracking behavior in flat-band multilayer graphene devices}

\author{Kry\v{s}tof Kol\'a\v{r}}
\email{Correspondence: kolar@zedat.fu-berlin.de}
\affiliation{Dahlem Center for Complex Quantum Systems and Fachbereich Physik, Freie Universit{\"a}t Berlin, 14195 Berlin, Germany}
\affiliation{Department of Applied Physics, Aalto University School of Science, FI-00076 Aalto, Finland}

\author{Dacen Waters}
\affiliation{Department of Physics, University of Washington, Seattle, Washington, 98195, USA}
\affiliation{Department of Physics and Astronomy, University of Denver, Denver, Colorado, 80210, USA}

\author{Joshua Folk}
\affiliation{Quantum Matter Institute, University of British Columbia,
Vancouver, British Columbia, V6T 1Z4, Canada}
\affiliation{Department of Physics and Astronomy, University of British Columbia,
Vancouver, British Columbia, V6T 1Z1, Canada}

\author{Matthew Yankowitz}
\affiliation{Department of Physics, University of Washington, Seattle, Washington, 98195, USA}
\affiliation{Department of Materials Science and Engineering,
University of Washington, Seattle, Washington, 98195, USA}

\author{Cyprian Lewandowski}
\address{National High Magnetic Field Laboratory, Tallahassee, Florida, 32310, USA}
\address{Department of Physics, Florida State University, Tallahassee, Florida 32306, USA}

\begin{abstract}
A central feature of many van der Waals (vdW) materials is the ability to
precisely control their charge doping, $n$, and electric displacement field,
$D$, using top and bottom gates. For devices composed of only a few layers, it
is commonly assumed that $D$ causes the layer-by-layer potential to drop linearly across the structure. Here, we show that this assumption fails
for a broad class of crystalline and moir\'e vdW structures based on Bernal- or
rhombohedral-stacked multilayer graphene. We find that the electronic
properties at the Fermi level are largely dictated by special layer--polarized
states arising at Bernal-stacked crystal faces, which typically coexist in the same band with layer-delocalized states.  We uncover a novel mechanism by which the layer-delocalized states completely screen the layer--polarized states from the bias applied to the remote gate. This screening mechanism leads to an unusual scenario where voltages on either gate dope the band as expected, yet the band dispersion and associated electronic properties remain primarily (and sometimes exclusively) governed by the gate closer to the layer--polarized states. 
Our results reveal a novel electronic mechanism underlying the atypical single-gate–-controlled transport characteristics observed across many
flat-band graphitic structures, and provide key theoretical insights essential
for accurately modeling these systems.  \end{abstract}
        
\maketitle

\def\layerproj{\mathcal{P}_l}
\def\dos{\rho}
\def\dosweight{n_{\text{K}}}
\def\dens{n}

\def\dstar{d^*}
\def\surface{layer--polarized}
\def\nsurf{N_{\text{pol}}}
\def\deltad{\Delta d}
\def\dstarmat{\boldsymbol{d}^*}
\def\deltadmat{\Delta \boldsymbol{D}}
\def\rhomat{\boldsymbol{\rho}}

\section{Introduction}

Dual-gated two-dimensional (2D) van der Waals (vdW) device structures offer unprecedented tunability, enabling simultaneous \textit{in situ} control of the charge density and perpendicular displacement field. This capability opens a broad experimental phase space, which has driven numerous recent discoveries in various multilayer vdW systems---particularly those featuring flat electronic dispersions~\cite{jarillo-herreroCaoCorrelatedInsulatorBehaviour2018,jarillo-herreroCaoUnconventionalSuperconductivityMagicangle2018,deanYankowitzTuningSuperconductivityTwisted2019,kimHaoElectricFieldTunable2021,yazdaniOhEvidenceUnconventionalSuperconductivity2021,efetovLuSuperconductorsOrbitalMagnets2019,jarillo-herreroCaoNematicityCompetingOrders2021,liLiuTuningElectronCorrelation2021,nadj-pergeAroraSuperconductivityMetallicTwisted2020,efetovStepanovUntyingInsulatingSuperconducting2020,youngSaitoIndependentSuperconductorsCorrelated2020,xuCaiSignaturesFractionalQuantum2023,xuParkObservationFractionallyQuantized2023,shanZengThermodynamicEvidenceFractional2023,liXuObservationIntegerFractional2023,feldmanFouttyMappingTwisttunedMultiband2024,youngZhouIsospinMagnetismSpintriplet2021,youngZhouHalfQuartermetalsRhombohedral2021,ashooridelaBarreraCascadeIsospinPhase2021,nadj-pergeZhangSpinOrbitEnhancedSuperconductivity2022,youngZhouSuperconductivityRhombohedralTrilayer2021,youngZhouIsospinMagnetismSpintriplet2021, juLuFractionalQuantumAnomalous2024,juLuExtendedQuantumAnomalous2024,yankowitzWatersInterplayElectronicCrystals2024, kimLiuTunableSpinpolarizedCorrelated2020,folkKuiriSpontaneousTimereversalSymmetry2022,jarillo-herreroCaoTunableCorrelatedStates2020,yankowitzWatersTopologicalFlatBands2024,yankowitzHeSymmetryBreakingTwisted2021,tutucBurgCorrelatedInsulatingStates2019,youngPolshynTopologicalChargeDensity2022, folkSuGeneralizedAnomalousHall2024}. Here, we focus on a striking and nearly universal feature that has been apparent since the earliest measurements of dual-gated moir\'e graphene multilayers, but that has received little attention until now: the frequent appearance of diagonal lines in the filling--displacement-field ($\nu-D$) plane.

\begin{figure*}[t]
    \centering
    \includegraphics[width=0.9\linewidth]
    {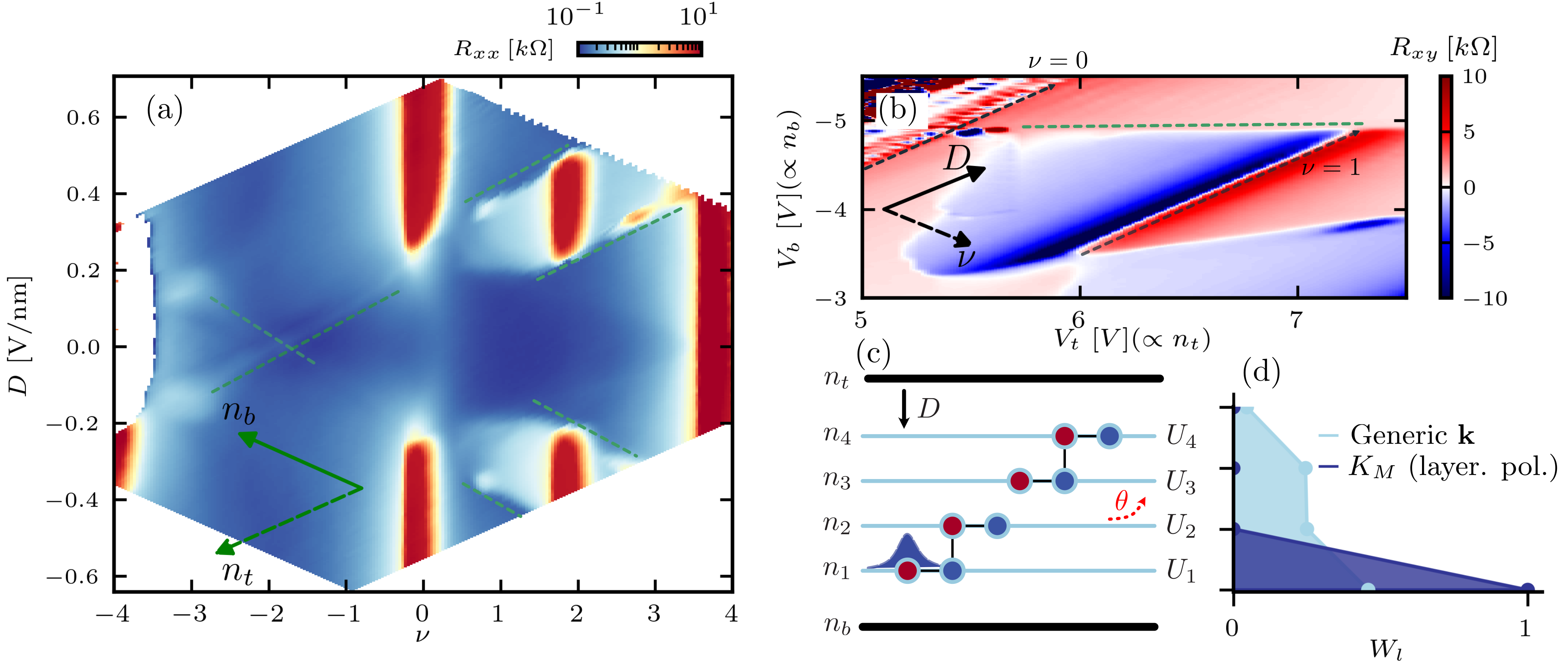}
    \caption{    
  (a)  Experimental $R_{ x x}$ map of twisted double bilayer graphene with twist angle $\theta=1.30^\circ$. The bottom and top gate axes are indicated by the green arrows. The onset of symmetry breaking at $\nu>0$ is tuned by $\dens_b$ for $D > 0$ and $\dens_t$ for $D < 0$. The data are from Ref.~\cite{yankowitzHeSymmetryBreakingTwisted2021}. Features closely tracking a single gate are highlighted with green dashed lines. 
   (b) Experimental $R_{ x y}$ map of twisted bilayer-trilayer graphene with twist angle $\theta=1.50^\circ$ as a function of the top ($V_t$) and bottom ($V_b$) gate voltages. Insulators at $\nu=0,\nu=1$ appear diagonally in this plot, and are highlighted using black dashed lines. The horizontal dashed green line denotes a transition between a normal metal and a quarter-metal phase. The data are from Ref.~\cite{yankowitzWatersTopologicalFlatBands2024}.
   (c) Schematic of a doubly-gated Bernal-terminated device (TDBG), illustrating the bottom layer sublattice-polarized surface state in blue. The blue lines denote graphene layers, and the black lines denote the top and bottom gate surfaces.
   (d) Relative layer occupations $W_l$ for the \surface{} state at $K_M$ (dark blue), and those of some generic bulk state (light blue). Note the perfect polarization of the \surface{} state to the bottom layer.}
    \label{fig:figone}
\end{figure*}

An example of such behavior is shown in Fig.~\ref{fig:figone}a for twisted
double bilayer graphene (TDBG), where diagonal features tracking a single gate
voltage have been observed in dozens of
devices~\cite{yankowitzWatersTopologicalFlatBands2024,kimLiuTunableSpinpolarizedCorrelated2020,folkKuiriSpontaneousTimereversalSymmetry2022,jarillo-herreroCaoTunableCorrelatedStates2020,yankowitzWatersTopologicalFlatBands2024,yankowitzHeSymmetryBreakingTwisted2021,tutucBurgCorrelatedInsulatingStates2019}.
The figure shows a longitudinal resistance ($R_{xx})$ map in the $\nu-D$ plane
for one such TDBG device, with arrows highlighting directions of constant
bottom ($n_b$) and top ($n_t$) gate charge density (proportional to $V_b$ and
$V_t$). Two key observations are a set of resistive bumps forming a cross-like
feature for hole-type doping ($\nu<0$) centered at $D=0$, and a high-resistance
region associated with broken spin and valley degeneracies for electron-type
doping ($\nu>0$) at larger $|D|$. The cross-like feature is approximately
aligned with axes of constant $n_b$ or $n_t$, as is the boundary of the
symmetry-breaking region.

The association of spin and valley symmetry breaking, hereafter referred to as flavour symmetry breaking, with a single gate voltage is even more evident in the transverse resistance ($R_{xy}$) map of Fig.~\ref{fig:figone}b, corresponding to a twisted bilayer-trilayer graphene device. The abrupt sign change in $R_{xy}$ appearing at $V_{b}=-4.9$~V corresponds to a transition from an unpolarized metallic phase to a metallic phase with full isospin degeneracy breaking. The symmetry-breaking transition is completely decoupled from $V_{t}$ over a wide voltage range. In contrast, other features of the maps in Figs.~\ref{fig:figone}a,b---such as band insulators and correlated insulators at integer $\nu$---exhibit no gate-tracking effect, remaining vertical in the $\nu-D$ plane and thus pinned to fixed $\nu$. In this context, the gate-tracking behavior is striking: it challenges the usual assumption that $D$ and $\nu$ alone define the boundaries of characteristic regions in experimental phase diagrams, suggesting instead that a single gate voltage can play a decisive role.

The 2D vdW materials that exhibit this gate-tracking behavior most prominently are TDBG~\cite{kimLiuTunableSpinpolarizedCorrelated2020,folkKuiriSpontaneousTimereversalSymmetry2022,jarillo-herreroCaoTunableCorrelatedStates2020,yankowitzHeSymmetryBreakingTwisted2021,tutucBurgCorrelatedInsulatingStates2019}, related structures with different numbers of twisted graphene layers~\cite{yankowitzWatersTopologicalFlatBands2024,folkSuGeneralizedAnomalousHall2024,shiXuTunableVanHove2021}, 
and rhombohedral graphene aligned with
hBN~\cite{juLuFractionalQuantumAnomalous2024,juLuExtendedQuantumAnomalous2024,yankowitzWatersInterplayElectronicCrystals2024}. A key microscopic feature of these graphene-based
systems is the presence of robust layer- and sublattice-polarized states at
the K and K' points of the monolayer Brillouin zone, arising from the local AB (Bernal) stacking arrangement between neighboring graphene sheets away from any twisted interface. The schematic in
Fig.~\ref{fig:figone}c shows the case of TDBG, formed by twisting two Bernal bilayers. 
In this system, there is a \surface{} state in the two outermost layers, but in other multilayer graphene systems the
relevant state may be internal \cite{yankowitzWatersTopologicalFlatBands2024}.
The bottom \surface{} state is relevant for $D>0$, making up a layer-polarized ``pocket" that coexists with more
delocalized states within a single band (Fig.~\ref{fig:figone}d). The gate-tracking behavior then generally arises from a combination of two effects: (i) the
layer-polarized pocket (on layer 1 in Fig.~\ref{fig:figone}c) predominantly controls the onset of symmetry-breaking phases due to its high density of states, and (ii) the delocalized states screen the layer-polarized pocket from the potential applied to the remote gate (the top gate in Fig.~\ref{fig:figone}c). 
The interplay of these two effects naturally leads to single-gate tracking of the symmetry-breaking boundary, as seen in
Figs.~\ref{fig:figone}a and \ref{fig:figone}b. 

In this paper, we analyze the gate-tracking mechanism to delineate its
microscopic origins, examine its ubiquity in moir\'e graphene structures, and
assess its robustness. We begin by clarifying the pivotal role of
layer-polarized states in shaping the band structure of Bernal-terminated
multilayer systems. Then, we introduce a simple electrostatic model that
provides an analytical view of how the \surface{} states evolve in the $\nu-D$
plane, revealing a novel mechanism by which the delocalized states screen the \surface{} states. Finally, we
apply this framework to TDBG, performing numerical mean-field simulations that may then be compared to experiment observations, for example in
Fig.~\ref{fig:figone}a. 
Although we focus on TDBG for clarity, our theory establishes a general mechanism that
applies to any multilayer systems with Bernal stacking as a component of its
structure, including rhombohedral multilayer graphene and twisted
bilayer-trilayer graphene. Appropriately generalized, our theory should also apply to any layered system featuring perfectly polarized states, such as twisted bilayer transition
metal dichalcogenides.

\begin{figure*}[t]
    \includegraphics[width=\linewidth]{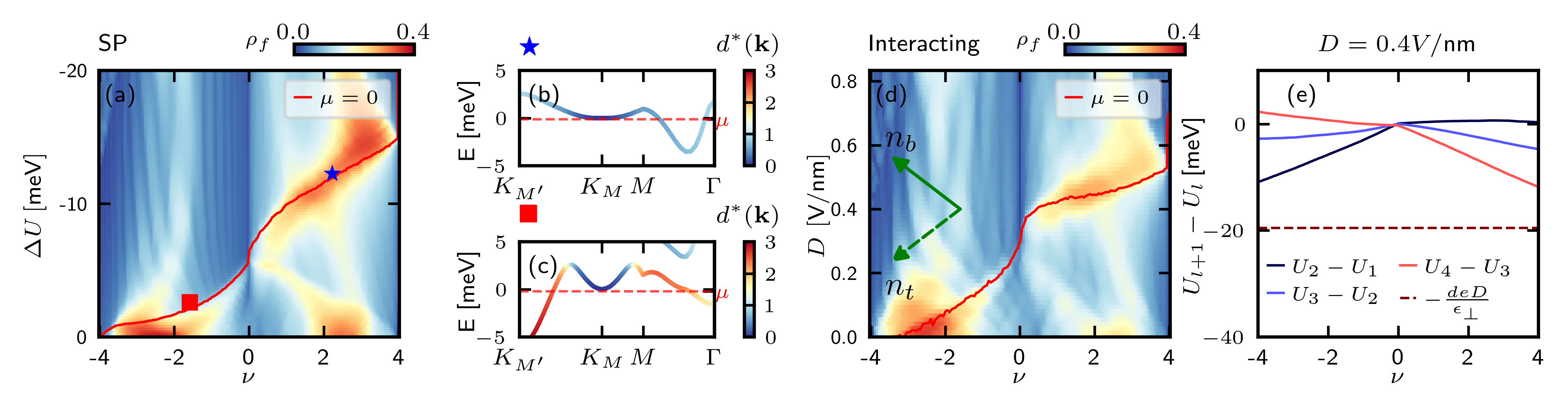}
    \captionof{figure}{(a) Single-particle (SP) $\Delta U$ vs $\nu$ map of the density of states per flavor, $\rho_f$. The contour of the \surface{} states from the Bernal stacking termination ($\mu=0$) is shown in orange.
    (b) Single-particle band structure at $\nu>0$ at the position of the blue star in (a). Color code shows the average distance of the state $d^*(\mathbf k)$ from the bottommost layer. 
    The bottom layer \surface{} states have $d^*(K_{M})=0$.
    (c) Same as (b), but at the position of the red square of (a).
    (d) Numerical $\nu-D$ colormap of the density of states per flavor ($\rho_f$) together with line of $\mu=0$ \text{meV}.
    We work in the Hartree approximation with dielectric constant $\epsilon_\perp=\epsilon_\parallel = 6$ (see Supplemental Materials~\cite{supplement} for details).
    (e) $U_{l+1}-U_l$ as a function of $\nu$ at $D=\SI{0.4}{V/nm}$ within the self-consistent Hartree calculation (obtained from Eq.~\eqref{eq:potdifferenceintermsofd}), compared to the naive value $-deD/\epsilon_\perp$.}

    \label{fig:figtwo}
\end{figure*}

\section{Results}

\subsection{Systems with a Bernal-stacked interface: perfectly layer-polarized states near the K point}

We begin by reviewing the properties of 2D graphene multilayer systems that feature a Bernal stacked interface. Although our discussion focuses on a bottom Bernal termination, the same ideas extend to bands residing primarily on internal Bernal-stacked interfaces (see Supplementary Materials \cite{supplement}). Labeling the atomic orbitals in the two terminating Bernal layers as $A1$, $B1$, $A2$, and $B2$---where $A/B$ denotes sublattices and $l=1,2$ labels the layers---local Bernal stacking implies that only the $B1 \to A2$ interlayer tunneling is significant. 
This arrangement yields a  state at the K point that is fully polarized to the bottom layer, even when other states away from the K point are not bound to the surface.

In the basis $\Psi_{\mathbf k} = (c_{\mathbf k,A1}, c_{\mathbf k,B1}, c_{\mathbf k,A2}, c_{\mathbf k,B2}, \dots)$, 
the single-particle Hamiltonian near the K point, for a device with Bernal stacking on the first two layers, takes the form
\cite{falkoMcCannLandauLevelDegeneracyQuantum2006,peresNilssonElectronicPropertiesBilayer2008,koshinoMcCannElectronicPropertiesBilayer2013}
\begin{equation}
\hsp + H_\perp=
\label{eq:hambernal}
\sum_{\mathbf k}
\Psi^\dagger_{\mathbf k}
    \begin{pmatrix}
        U_1 &  v_F \overline{k}&-v_4 \overline{k}&-v_3 k &0& \\
         v_F k& U_1&t_1  &-v_4 \overline{k} &\\ 
         -v_4 k& t_1&U_2 &  v_F \overline{k}& \\
        -v_3  \overline{k}&-v_4 k& v_F k &U_2&  \\
        0& &&& \ddots\\
        \vdots&  
    \end{pmatrix}
\Psi_{\mathbf k},
\end{equation}
where $k=k_x+ik_y$, $\overline{k}=k_x - ik_y$ momenta are measured from the K point, $v_F$ is the graphene Dirac velocity,
and $v_3, v_4 \ll v_F$ denote nonlocal interlayer tunneling velocities, and $t_1$ is the strength of $B1\to A2$ tunneling.
A key ingredient in the above Hamiltonian is the layer-dependent electrostatic potential $U_l$ (depicted in Fig. \ref{fig:figone}c). In the next section, we derive how $U_l$ depends on charges on the top ($n_t$) and bottom ($n_b$) gates. 

The \surface{} state on the $A1$ orbital is an exact layer and sublattice polarized eigenstate of Eq.~\eqref{eq:hambernal} at the K point ($\mathbf k=0$) with energy $U_1$ and is also spin- and valley-degenerate.
For small $\mathbf k$ away from the K point, states retain strong layer polarization, forming a well-defined pocket of \surface{} states. 
The peculiarity of moir\'e systems featuring Bernal interfaces that distinguishes them from standard Bernal bilayer graphene is that this pocket exists within a well-defined flat moir\'e band. As we will show, this high density-of-states pocket controls symmetry breaking, and responds primarily to the proximal gate as a result of its layer polarization. For simplicity, throughout the remainder of the paper we measure energies with respect to $U_1$, effectively defining $U_1=0$. In this convention, the \surface{} state is at the Fermi level when the chemical potential is zero ($\mu=0$).

To illustrate the role of the \surface{} pocket, we now examine its impact on the band structure of TDBG. Fig.~\ref{fig:figtwo}a shows a map of the single-particle density of states (DOS) calculated as a function of the filling factor $\nu$ and an interlayer potential difference $\Delta U=U_{l+1}-U_l$ (see the Supplemental Material \cite{supplement} for full details of the model). Here, $\nu$ and $\Delta U$ are treated as theoretical parameters; later, we will connect them to experimentally tunable gate charges. A prominent feature in Fig.~\ref{fig:figtwo}a is a region of enhanced DOS associated with a flat part of the band at positive $\nu$. Overlaying the $\mu = 0$ contour reveals that this high-DOS region coincides with the \surface{} state being exactly at the Fermi level. A similar high DOS feature also appears for negative filling, although it aligns less precisely with the $\mu=0$ condition. 

The interpretation of the \surface{} pocket as the source of the high density of states 
is confirmed by considering the band structures at specific points along the $\mu=0$ contour, shown in  Figs.~\ref{fig:figtwo}b and \ref{fig:figtwo}c,
where we color-code each state in the band by its average distancefrom the bottom layer,
$\dstar(\mathbf k)\equiv \sum_{l}(l-1) W_l(\mathbf k)$, where $W_l(\mathbf k)$
is the layer distribution. The  pocket is centered around the K point of the
lower bilayer, $K_M$.  For $\nu>0$ (Fig.~\ref{fig:figtwo}b) the entire band
(not just the pocket) is strongly polarized to the bottom of the structure,
resulting in a quenching of the \surface{}
pocket dispersion. Fig.~\ref{fig:figtwo}c shows the contrasting case of
$\nu<0$, in which most of the moir\'e band is widely distributed across the
layers leading
to a larger dispersion of the \surface{}
pocket. In what follows, we employ the zero-energy
\surface{} state (at $U_1=0$) as a theoretical reference point, serving to
clarify how the \surface{} pocket evolves with gating in the band structure and
explaining the experimental single-gate tracking.

\subsection{Relating layer potentials to the gate charges}

We now develop a framework for properly describing the electrostatics of multilayer devices, which is crucial for understanding the band structure evolution with density $\nu$ and displacement field $D$ (rather than the interlayer potential difference $\Delta U$). Consider a multilayer system with $N$ layers labeled $l=1,\ldots,N$, sandwiched between the bottom and top gates, as shown
in Fig.~\ref{fig:figone}c. We denote the net electron densities in each layer
$l$ by $\dens_l$, and in the top and bottom gates by $\dens_t$ and $\dens_b$,
respectively. These densities relate to the top and bottom gate voltages as
$\dens_t = -\frac{1}{e}C_{tg} V_t$ and $\dens_b = -\frac{1}{e}C_{bg} V_b$,
where $C_{tg}$ and $C_{bg}$ are the top and bottom gate capacitances per unit
area, and $e$ is the electron charge. Overall charge neutrality implies that
the sum of these gate charges fixes the total device density, $\dens = \sum_l
\dens_l = -(\dens_t + \dens_b)$. Their difference sets the experimentally
accessible displacement field, $D= e\frac{\dens_b-\dens_t}{2\epsilon_0} =
\frac{C_{tg} V_t - C_{bg} V_b}{2 \epsilon_0}$.  The full Hamiltonian is thus
\begin{equation}
\label{eq:ham}
\hat{H} = \hsp+H_{\text{int},\mathbf q \neq 0} +H_\perp,
\end{equation}
where $\hsp$ is the single-particle Hamiltonian at zero out-of-plane electric field ($U_l=0$), $H_{\text{int},\mathbf q \neq 0}$ is the finite-momentum density-density Coulomb interaction (excluding its uniform, multilayer component), and $H_{\perp}$ is the uniform, multilayer part of the Coulomb interaction originating from out-of-plane electric fields~\cite{lewandowskiKolarElectrostaticFate$N$layer2023,koshinoMcCannElectronicPropertiesBilayer2013,koshinoKoshinoInterlayerScreeningEffect2010,mccannKoshinoGateinducedInterlayerAsymmetry2009,guineaGuineaChargeDistributionScreening2007}:
\begin{equation}
\label{eq:hperp}
H_{\perp}=\sum_{l=1}^{N} U_l\,\, \layerproj
\end{equation}
where $\layerproj$ is the projector onto layer $l$,
and $U_l$ are the self-consistent layer potentials.
To find $U_l$, we integrate Gauss' law across the layers:
\begin{equation}
\label{eq:potdifference}
    U_{l+1}-U_{l} = -e^2 d \frac{\dens_b+ \sum_{j \leq l}\dens_j }{\epsilon_\perp\epsilon_0},
\end{equation}
with $d$ the interlayer distance and $\epsilon_\perp$ the out-of-plane dielectric constant. Together, $\hsp + H_\perp$ form the single-particle description of Eq.~\eqref{eq:hambernal}, with $U_l$ subject to the recursion relation of Eq.~\eqref{eq:potdifference}
\footnote{$H_\perp$ can also be obtained by considering the standard density-density interaction and incorporating the layer dependence of the Coulomb interaction. In that viewpoint, the $H_\perp$ terms arise from the $q=0$ Hartree contribution of charges localized in different layers~\cite{lewandowskiKolarElectrostaticFate$N$layer2023}. To build qualitative intuition for $H_\perp$, however, we focus here on Eq.~\eqref{eq:potdifference} for the layer potentials $U_l$.}.

This formulation treats out-of-plane electric fields due to charges in the gates and in the
sample on the same footing, in keeping with Gauss' law. The practice of incorporating
an unscreened displacement field $D =e\frac{\dens_b-\dens_t}{2\epsilon_0}$ in
the microscopic Hamiltonian of Eq.~\eqref{eq:hambernal} as a constant potential
difference $\Delta U= -d e D/\epsilon_\perp$ between layers, as is typically
done in single-particle approaches, is generally not correct. 
This is best seen by rewriting the potential difference from Eq.~\eqref{eq:potdifference} explicitly in terms of $D$,
\begin{equation}
\label{eq:potdifferenceintermsofd}
    U_{l+1}-U_{l} =-\frac{d e D}{\epsilon_\perp}  +\frac{e^2 d}{2\epsilon_\perp\epsilon_0}\left[n-2 \sum_{j \leq l}\dens_j\right].
\end{equation}
The second term, present even at zero displacement field, is necessary to comply with Gauss' law. 
Already at zero doping ($n=0$), this term enables screening of the displacement field by the system \cite{koshinoMcCannElectronicPropertiesBilayer2013,koshinoKoshinoInterlayerScreeningEffect2010,mccannKoshinoGateinducedInterlayerAsymmetry2009}.
For nonzero doping, it also includes the additional electric field from the gates due to doping \cite{lewandowskiKolarElectrostaticFate$N$layer2023}.
Capturing both effects is crucial for an accurate analysis in the $\nu-D$ plane.

To illustrate the merits of this framework,
Fig.~\ref{fig:figtwo}d shows the
self-consistently determined Hartree DOS, along with the contour of $\mu=0$ (e.g., the
\surface{} state energy). Note that here we treat the finite-$\mathbf q$ interaction
within the Hartree approximation, neglecting exchange effects and ignoring the possibility of isospin symmetry breaking.
Importantly, in contrast with Fig.~\ref{fig:figtwo}a, the vertical axis here is $D$.
Using Gauss' law then allows us to also plot the gate density ($\dens_t$,$\dens_b$) axes in green. 
We further note that our method avoids bias by using the same plane wave basis set for different values of $D$ and $\nu$ (see Supplementary Material \cite{supplement}).

A key observation is that, for $\nu<0$, the \surface{} state $\mu=0$
contour aligns with the $\dens_t$ axis (constant $\dens_b$). We will show in the following section that this \surface{} state behavior arises due to the large 
spatial separation between the \surface{} state in the bottom layer and
the rest of the highest moir\'e valence band, which is situated towards the
center of the TDBG structure (see Fig.~\ref{fig:figtwo}c) and efficiently screens the top gate.
For $\nu>0$, the \surface{} state contour is significantly more horizontal than the $\dens_t$ axis, which we will show results from overscreening.
Upon developing a physical understanding of the \surface{} state behavior in the two-gate geometry, we will show later that when symmetry breaking is included,
the system recovers the experimentally seen single gate tracking even at $\nu>0$. 

To further illustrate the importance of determining the layer-potentials correctly,
in Fig.~\ref{fig:figtwo}e we show $U_{l+1}-U_l$ within the self-consistent Hartree calculation compared to the ``conventional'' result of $U_{l+1}-U_l = -\frac{d e D}{\epsilon_\perp}$. In the conventional approach (i.e., keeping only the first term in Eq.~\eqref{eq:potdifferenceintermsofd}), the potentials do not depend on $\nu$, unlike the self-consistent solution (i.e., keeping all terms in Eq.~\eqref{eq:potdifferenceintermsofd}) which strongly depends on $\nu$. Crucially, the true potentials deviate from the conventional result not only in magnitude, but also in the sign of the energy difference, suggesting a possibility for non-trivial state renormalization with external displacement field. Lastly, in Supplementary Fig. \ref{fig:figsuppla} we show the expected $\mu=0$ contours computed using the self-consistent framework and the conventional approach for several dielectric constants $\epsilon_\perp$. This comparison highlights that only with layer-potentials single gate tracking occurs generically (for $\nu <0$) for a wide range of dielectric constants. A calculation using conventional relation $U_{l+1}-U_l$ with $D$ would only yield single gate tracking for a limited doping range if the dielectric constant is fine-tuned.

\def\rholoc{\rho}
\def\invrholoc{\frac{1}{\rho}}

\subsection{Physical origins of the gate tracking behavior} 
\label{sec:physical_model}

We now develop a qualitative understanding of the screening mechanisms at play in determining the evolution of the \surface{} state (or equivalently the $\mu = 0$ contour) in the $\nu-D$ plane, using an analytical model that includes the effects of both $\hsp$ and $H_\perp$ in Eq.~\eqref{eq:ham}. 
We first write down the compressibilities relative to the top and bottom gates, $\frac{\partial \mu}{\partial \dens_t}$ and $\frac{\partial \mu}{\partial \dens_b}$, given by
\begin{eqnarray}
\frac{\partial \mu}{\partial \dens_t} &=&  \frac{\partial \mu}{\partial \dens} \frac{\partial \dens}{\partial \dens_t} + \sum_{l} W_l  \frac{\partial U_l}{\partial \dens_t} \label{eq:somea}\\
    \frac{\partial \mu}{\partial \dens_b} &=&  \frac{\partial \mu}{\partial \dens} \frac{\partial \dens}{\partial \dens_b} + \sum_{l} W_l  \frac{\partial U_l}{\partial \dens_b}\label{eq:someb}
\end{eqnarray}
where $W_l = \partial \mu / \partial U_l$ denotes fraction of electronic states
in layer $l$ at the Fermi level ($\sum_l W_l = 1$)\footnote{Physically, this relation between $W_l $ and $\partial \mu /
\partial U_l$ can be understood by realizing that the total chemical potential
shifts due to the layer potentials with the amount of each shift set by the
relative density of charges at the Fermi level on that layer.}.

\begin{figure}[t]
    \includegraphics[width=\columnwidth]{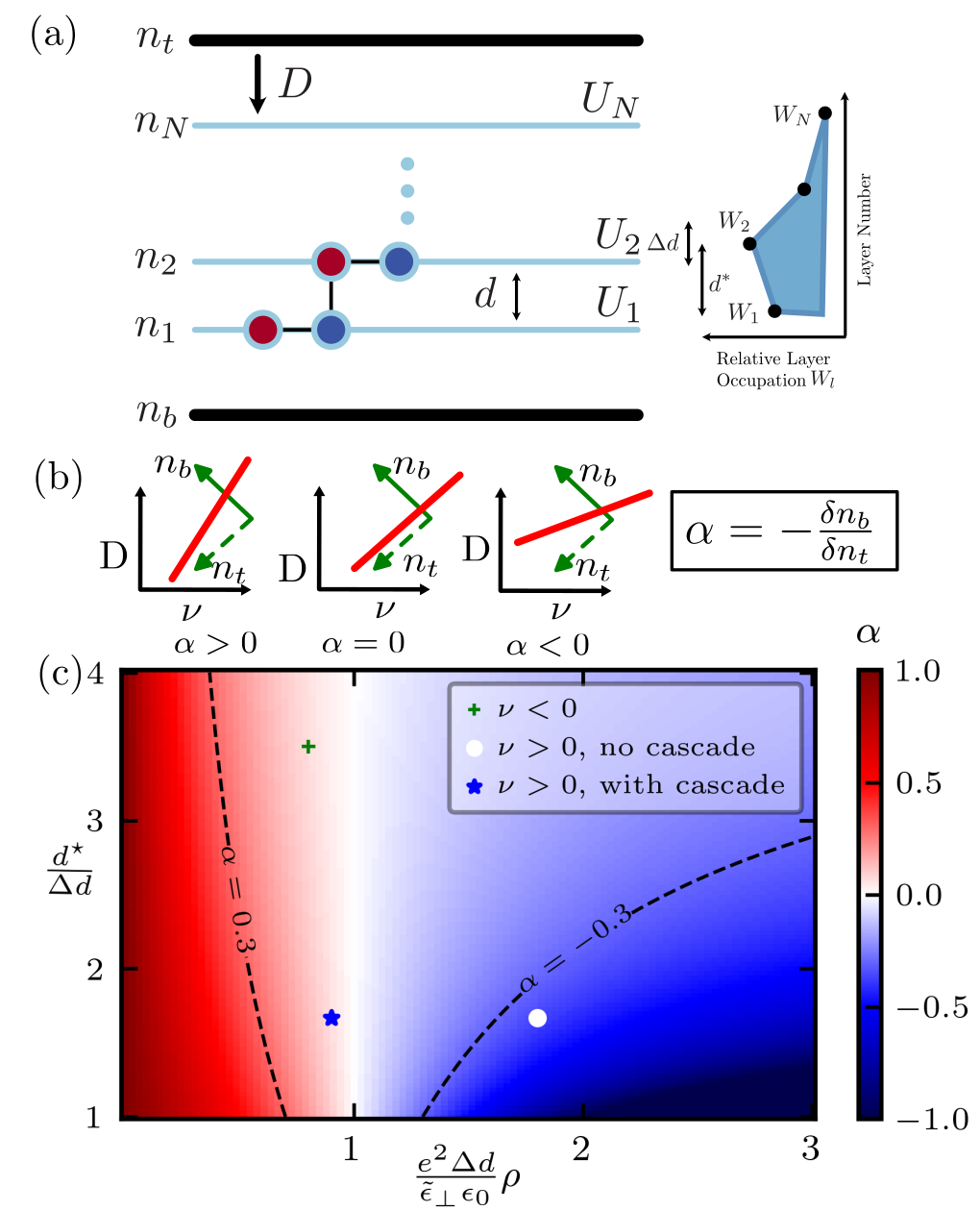}
    \caption{
    (a) Sketch of the theoretical setup, showing the gates and the layered material. We assume the bottom two layers have Bernal (AB) stacking, leading to \surface{} states at the bottommost surface. The average distance of carriers from the bottom layer $\dstar$ and the average vertical spread $\deltad$ are shown.
    (b) Schematic illustration of the behaviors expected for different values of $\alpha$.
    (c) Map of $\alpha$ (Eq.~\eqref{eq:defalpha}) as a function of $\dstar/\deltad$ and $\frac{e^2 \deltad}{\epsilon_\perp \tilde\epsilon_0}\rho$.
    $\alpha=0$ corresponds to perfect gate tracking. Contours of $|\alpha|=0.3$ are shown in dashed black. 
    Estimated parameter values for TDBG at $\nu<0$ and $\nu>0$ without and with flavor symmetry breaking (cascade) are also shown.}
    \label{fig:figthree}
\end{figure}

Due to the explicit presence of the $\dens_b$ term in Eq.~\ref{eq:potdifference}, 
the potentials $U_l$ respond differently to the two gates (see Supplemental Materials \cite{supplement} for derivation).
This leads to \begin{eqnarray}
-\frac{\partial \mu}{\partial \dens_b} &=&\invrholoc - \frac{e^2 \deltad}{\tilde \epsilon_\perp\epsilon_0}+\frac{e^2 \dstar }{\tilde \epsilon_\perp\epsilon_0}\label{eq:dmdnb} \\
-\frac{\partial \mu}{\partial \dens_t} &=&\invrholoc  - \frac{e^2 \deltad}
{\tilde \epsilon_\perp\epsilon_0}\label{eq:dmdnt}
\end{eqnarray}
where we relate $\partial \mu / \partial n$ with the inverse density of states, $1/\rho$, at the Fermi level, 
and use an effective dielectric constant $\tilde \epsilon_\perp> \epsilon_\perp$ to account for additional screening by bands away
from the Fermi level. We have also introduced two effective parameters, $\dstar$ and $\deltad$, defined via
\begin{eqnarray}
\dstar  &=&  d \sum_{l=1}^{N} (l-1) W_l\\
\label{eq:defdeltad}
\deltad &=& \frac{d}{2} \sum_{l,l'=1}^{N}\left|l'-l\right| W_l W_{l'}.
\end{eqnarray}
which capture, respectively, the average distance of the charges at the Fermi level from the bottommost ($l=1$) layer and their average vertical spread across the layers, as illustrated in Fig.~\ref{fig:figthree}a. Note that $\dstar \geq \deltad$, and both $\rholoc$ and $W_l$ generally depend on the instantaneous gate configurations. 

These expressions can be understood as follows. In addition to the standard
compressibility $\invrholoc$, the states at the Fermi level experience
electrostatic shifts due to changes in the layer potentials $U_l$ as the gate
charges vary. Both gates induce the same change in the layer distribution, $\delta
\dens_l$, yielding a self-interaction term proportional to $\deltad$.
Meanwhile, the bottom gate adds an extra shift proportional to $\dstar$, the
average distance from the bottom layer. This asymmetry arises
due to the explicit $\dens_b$ term in Eq.~\eqref{eq:potdifference}, which 
comes from the gauge choice $U_1 = 0$, pinning the \surface{} state energy to $\mu$ alone (see
discussion below Eq.~\eqref{eq:hambernal}). In a different gauge, the relevant
quantity for the \surface{} state would be $\mu - U_1$, but the same physical
conclusions would follow.

Next, we relate the gate-projected compressibilities in Eq.~\eqref{eq:dmdnt} to the experimentally observed \surface{} state trajectory in the $n_b, n_t$ plane. For small changes in gate charges, the chemical potential changes as
\begin{equation}
\mathrm{d} \mu= \frac{\partial\mu}{\partial \dens_b} \delta \dens_b +\frac{\partial \mu}{\partial \dens_t} \delta \dens_t\,,
\end{equation}
where $\delta \dens_{t/b}$ are small increments in the top/bottom gate densities. Constant-$\mu$ trajectories thus satisfy
 \begin{equation}
 \delta \dens_b = -\alpha \delta \dens_t,
 \end{equation}
where we introduce
\begin{equation}
\label{eq:defalpha}
\alpha \equiv \frac{\partial\mu}{\partial \dens_t}/  \frac{\partial\mu}{\partial \dens_b} = \frac{1 -\frac{e^2 \deltad}{\tilde\epsilon_\perp\epsilon_0}\rholoc}
{ 1 +(\frac{\dstar}{\deltad}-1)\frac{e^2 \deltad}{\tilde\epsilon_\perp\epsilon_0}\rholoc},
\end{equation}
as the ratio of the two gate-projected compressibilities defined in
Eq.~\eqref{eq:dmdnt}. The parameter $\alpha$ is directly connected to screening
of the \surface{} pocket from the top gate potential, setting the slope of the
constant-$\mu$ contours as sketched schematically in
Fig.~\ref{fig:figthree}b.  $\alpha\rightarrow 0$ corresponds to perfect
screening: the $\mu=0$ contour runs  parallel to the $\dens_t$ axis, implying
that the \surface{} state is  tuned only by the bottom gate. For $\alpha>0$,
the contour tilts toward the $D$ axis.  The \surface{} pocket is underscreened from the top gate, and at
$\alpha=1$  it is tuned equally by both gates following the naive picture often
applied to 2D stacks. For $\alpha<0$, the contour tilts
towards the $\nu$ axis, corresponding to overscreening. 
We note at this point that in Eq.~\eqref{eq:defalpha}, the parameters should be mathematically understood
to correspond to the screening (delocalized) states only and excluding the contribution of the \surface{} pocket. 
We derive this result in the Supplement \cite{supplement}, but intuitively, it arises because along the $\mu=0$ contour,
the \surface{} states occupation is constant, and they do not contribute to screening.

To gain further insight, we plot $\alpha$ as a function of the dimensionless
parameters $\frac{\dstar}{\deltad} \geq 1$ and $\frac{e^2
\deltad}{\tilde\epsilon_\perp\epsilon_0}\rholoc$ in Fig.~\ref{fig:figthree}c.
$\frac{\dstar}{\deltad}$ reflects the degree to which the layer-polarized and delocalized states can be thought of as independent conducting sheets, whereas 
$\frac{e^2\deltad}{\tilde\epsilon_\perp\epsilon_0}\rholoc$
reflects the strength of out-of-plane Coulomb
self-interactions with respect to kinetic energy.  
When there are no screening states, $\rholoc=0$ so $\alpha=1$, and $\mu$ depends only on filling $\nu$.
Increasing the number of screening states, $\rholoc$,
causes $\alpha$ to decrease until
\begin{equation}
\label{eq:conditionrho}
\frac{e^2 \deltad}{\tilde \epsilon_\perp\epsilon_0}\rholoc = 1,
\end{equation}
at which point $\alpha=0$ and the top gate is completely screened
($\frac{\partial \mu}{\partial \dens_t} =0$). For even larger $\rholoc$, $\alpha<0$, which corresponds to overscreening. Increasing the ratio $\frac{\dstar}{\deltad}$ serves to broaden the regime of $|\alpha|\ll 1$ (see, for example, the dashed curve corresponding to $|\alpha|=0.3$ in Fig.~\ref{fig:figthree}c). This corresponds to a situation in which the delocalized states have a larger vertical separation from the layer-polarized states, and can thus screen the adjacent gate more effectively.

We now comment on three relevant theoretical extensions of this model, all of which act to increase $\alpha$ (i.e., move system parameters effectively to the left in the Fig.~\ref{fig:figthree}c). First, the in-plane Hartree correction renormalizes the bandstructure. However, the \surface{} state of Eq.~\eqref{eq:hambernal} has zero in-plane inhomogeneity, and is not (to first order) affected by in-plane Hartree.  As a result, the non-surface states will shift upward with respect to the \surface{} state, reducing the effective value of $\rholoc$ and increasing $\alpha$. Second, if the states of
interest in the \surface{} pocket have some finite weight on the second layer
$W^{lp}_2$, then they are at the Fermi level when $\mu-
W^{lp}_2U_2 =0$. We show in the Supplement \cite{supplement}
that this effect also increases $\alpha$.
Third, due to its strong layer localization, we expect exchange effects to be stronger within the \surface{} pocket than
within the rest of the band. Effectively, upon doping, the screening states shift upward relative to the \surface{} pocket,
decreasing the effective $\rholoc$, and increasing $\alpha$.

\begin{figure}[t]
\includegraphics[width=\columnwidth]{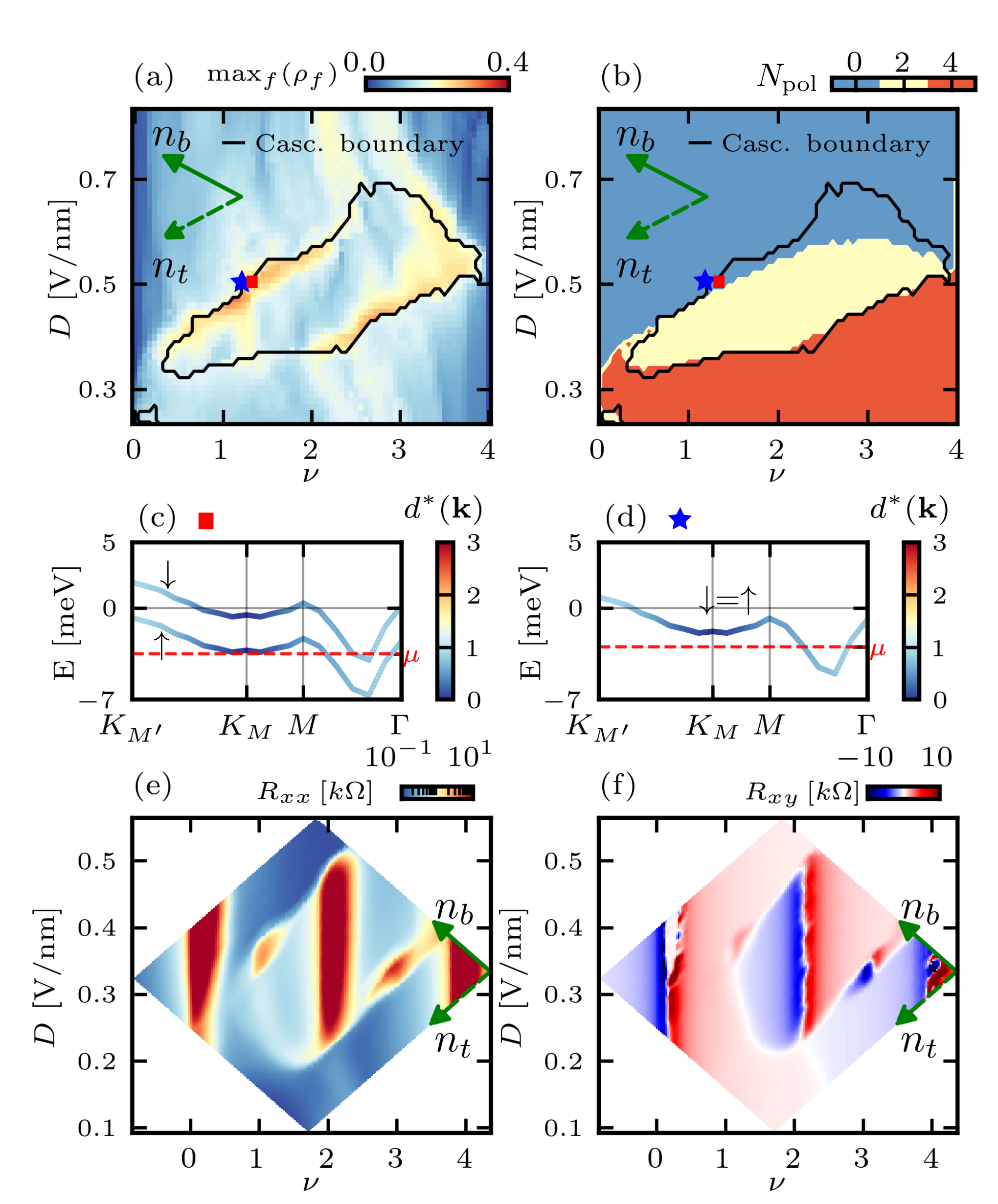}
    \caption{ 
  (a) Density of states for the dominant flavor for $\intstrengthfock=\SI{5.9}{meV}$ and $\epsilon=\epsilon_\perp=6$, with the boundary of cascade region highlighted.
  (b) Same as (a) but for \surface{} state occupation $\nsurf$. 
    (b) Spin-resolved band structures at the red square point of (a), inside the symmetry broken region.
    (c) Spin-resolved band structures at the red square point of (a), inside the symmetry broken region.
    (d) Spin-resolved band structures at the blue star point of (a), just outside the symmetry breaking region. Note the bands are spin degenerate.
    (e) Experimental $R_{ x x}$ map of twisted double bilayer graphene, zoomed in to
    the symmetry breaking region. The data are from Ref.~\cite{yankowitzHeSymmetryBreakingTwisted2021}.
    (f) Same as (e) but for $R_{xy}$.}
    \label{fig:figfour}
\end{figure}

We conclude this section by highlighting a subtle and surprising implication of Eq.~\eqref{eq:conditionrho}, which is that single-gate tracking can occur even in few-layered systems.  The microscopic origin is the vertical delocalization ($\deltad \neq 0$) of the screening states. An intriguing feature of this mechanism is that it allows overscreening ($\alpha<0$) to occur.
This contrasts with more trivial effects leading to single gate tracking in thick rhombohedral graphite
\cite{mishchenkoShiElectronicPhaseSeparation2020}. In this
system, electronic states are highly localized ($\deltad \approx 0$) to the outermost surface surface layers of the crystal. Here, Eq.~\eqref{eq:defalpha} gives $\alpha \to \left[1 + \frac{e^2 \dstar}{\tilde
\epsilon_\perp\epsilon_0}\rholoc \right]^{-1}$, which always satisfies $\alpha >0$. Good gate tracking (small
$\alpha$) will then occur only in a setting where the offset between the two outermost layers ($\dstar$) is very large---a condition
naturally satisfied in rhombohedral graphite given that the low energy states reside on the surfaces.

\subsection{Relation to experimental observations}

We now apply the above mechanism to real systems and connect to experimental observations. First, consider TDBG at $\nu < 0$, where we found that the \surface{} state in Fig.~\ref{fig:figtwo}d
was primarily tuned by a single gate. In this regime of TDBG, $\dstar \approx 2d$, while $\deltad\approx0.6 d$, since
the screening states (see Fig.~\ref{fig:figtwo}c) are concentrated near the other end of the stack.
Fig.~\ref{fig:figthree}c then predicts a robust large region of small $|\alpha|$ in parameter space, which we confirm in the Supplementary \cite{supplement}, finding 
the slope of the $\mu=0$ contour to be largely independent of the dielectric constant $\epsilon_\perp$.
Coming back to the experimental result in Fig.~\ref{fig:figone}a, we interpret the gate-tracking cross-like features for $\nu < 0$ as corresponding to the \surface{} state
crossing the Fermi level.

Next, we consider the gate-tracking of symmetry-breaking boundaries occurring at $\nu>0$. Although the precise value of $\alpha$ varies between systems, it appears to be an experimental fact that $\alpha>0$ for most symmetry-breaking phase boundaries.
On the other hand, in TDBG in the flavor-degenerate calculation of Fig.~\ref{fig:figtwo}d, the \surface{} state is overscreened ($\alpha<0$),
a behavior enabled by the delocalization of the screening states. However, since the $\mu=0$ contour is associated with a high density of states,  
we expect the system, according to the Stoner criterion \footnote{The regime of $\alpha < 0$ occurs when $\rho e^2 \deltad / \tilde{\epsilon}_\perp \epsilon_0 \gg 1$. By comparison the Stoner criterion takes the form $\rholoc\frac{e^2}{\tilde{\epsilon} \epsilon_0 k_F} = 1$, where $k_F$ is the Fermi momentum and $\tilde{\epsilon} \simeq
 \tilde{\epsilon}_\perp$ is the relevant (likely in-plane) dielectric constant.
 Generically, 
 $1/k_F \gtrsim \deltad$ as average inter-particle separation exceeds the layer
 spacing. As such, the Stoner criterion will typically be satisfied before $\rho e^2 \deltad / \tilde{\epsilon}_\perp
 \epsilon_0 \gg 1$. The system will then prefer to undergo symmetry breaking and reduce
 $\alpha$.}, to be unstable towards symmetry breaking, reducing the density of states.
Provided that phase transitions are triggered when the \surface{} state crosses the Fermi level within the symmetry broken phase,
the evolution of the phase boundary in the $\nu-D$ plane follows from our present framework.
Crucially, reduction of the density of states effectively moves the system parameters towards the left of Fig.~\ref{fig:figthree}c, increasing the value of $\alpha$. At $\nu>0$, TDBG has
has $\dstar\approx d$ and $\deltad\approx 0.6d$. A back of the envelope calculation, estimating $\rholoc \approx 0.1 N_f/\Auc$ $(\text{meV})^{-1}$, where  $N_f$ is the number
of active flavors, gives $\frac{e^2 \deltad}{\tilde \epsilon_\perp\epsilon_0}\rholoc \approx  0.44 \cdot N_f$
for $\tilde \epsilon_\perp \approx 8$.
In an unpolarized phase, $N_f=4$, predicting overscreening with $\alpha<0$. If instead symmetry-breaking occurred (i.e., a half-metal phase with $N_f=2$) a small positive $\alpha$ would be obtained, as depicted in Fig.~\ref{fig:figthree}c.

We now show that the \surface{} state is indeed relevant for symmetry breaking
transitions.
To that end, we adopt a minimal mean-field model of flavor symmetry breaking~\cite{ilaniZondinerCascadePhaseTransitions2020}, which shifts each flavor by an amount proportional to its filling
\begin{equation}
\label{eq:defhfock}
H_{Fock}  = -\intstrengthfock \diag(\nu_{K\uparrow},\nu_{K'\uparrow},\nu_{K'\downarrow},\nu_{K\downarrow}),
\end{equation}
where $\nu_f$ is the filling of flavor $f$ relative to charge neutrality, and $\intstrengthfock$ sets the interaction strength. For simplicity, we only consider valley-degenerate symmetry breaking by imposing $\nu_{K'\uparrow}= \nu_{K\uparrow}$ and $\nu_{K'\downarrow}= \nu_{K\downarrow}$---meaning the only competing phases at noninteger fillings
are a spin-polarized half-metal, and metal with unbroken flavor symmetry.

By optimizing the charge distribution between the two spins for different values of $\nu$ and $D$, we obtain the phase diagram shown in Fig.~\ref{fig:figfour}a (see Supplementary material \cite{supplement} for details of the calculation).
The boundary of the symmetry-broken region, shown with a black line, is marked by a sharp drop in the density of states.
Furthermore, close to the symmetry-breaking boundaries, the \surface{} state occupation $\nsurf$ (i.e., how many of the 4 spin/valley \surface{} $\mathbf k=0$ states are occupied) changes nearly simultaneously (as shown in Fig.~\ref{fig:figfour}b).
These plots support an interpretation of the transition from the half-metal phase into the symmetry unbroken state as arising 
from the depletion of the \surface{} pocket, which has high density of states.
This interpretation is further confirmed by inspecting the band structures close to the transition point, shown in Figs.~\ref{fig:figfour}c,d.
In the polarized (half-metal) phase (Fig.~\ref{fig:figfour}c), the high density of states surface pocket is emptied just before the transition. This emptying of the surface pocket renders the polarized state energetically unfavorable, triggering a transition into the unpolarized state (Fig.~\ref{fig:figfour}d). 
The slope of the transition line is thus set by the slope of the \surface{} pocket within the symmetry broken phase.
This calculation captures the experimental observations from Fig.~\ref{fig:figone}a, along with the high-resolution zoom-in maps from the TDBG device, which
we show in Figs.~\ref{fig:figfour}e,f. The present pocket-depletion mechanism applies for transitions from a polarized into a symmetry-unbroken phase, 
such as the half-metal to metal transition in TDBG (Fig.~\ref{fig:figfour}f), and the
quarter-metal to metal transition in twisted bilayer-trilayer graphene (Fig.~\ref{fig:figone}b). 

As mentioned above, $\alpha$ is typically positive, but varies from system to system and as a function of twist angle. For example, the half-metal to metal transition in twisted monolayer-bilayer graphene \cite{shiXuTunableVanHove2021,yankowitzWatersTopologicalFlatBands2024} has a larger $\alpha$
than for TDBG (Fig.~\ref{fig:figfour}f). Within our framework, this would naturally occur because of a smaller delocalization and/or density of states of the screening states.
On the other hand, in twisted bilayer-trilayer graphene, the transition from the quarter-metal exhibits perfect gate tracking (Fig.~\ref{fig:figone}b),
which would imply higher density of states than TDBG. This conclusion is further supported by the observation of spontaneous electronic crystallization behavior at $\nu=1/4$~\cite{folkSuGeneralizedAnomalousHall2024}. 
Lastly, we comment on the apparent lack of experimental manifestation of features with $\alpha<0$ slope. 
The layer dependent Coulomb interaction~\cite{lewandowskiKolarElectrostaticFate$N$layer2023} in the Fock term, as well as the effect of finite delocalization of the \surface{} pocket would act to increase the value of $\alpha$, making experimental observation of $\alpha<0$ less likely, but still possible in principle.

\section{Discussion}

Motivated by unexplained experimental trends, we have identified a
largely overlooked element of theoretical modeling in both crystalline and
moir\'e graphene multilayer structures: the self-consistently determined layer potentials. Our findings broadly apply to systems with flat electronic bands and a Bernal-stacked interface. 
By analyzing the microscopic Hamiltonian, we showed that perfectly
\surface{} states due to Bernal-stacked interfaces appear as a pocket in
momentum-space, forming a portion of a larger moir\'e band.
We further showed that this pocket is responsible for symmetry breaking transitions and other features in the $\nu-D$ plane. 
Exploiting its perfect layer polarization, we constructed
an analytic framework based on accurate electrostatics to derive 
necessary criteria for the single-gate tracking of the \surface{} pocket, uncovering
a novel mechanism by which it is screened from a remote gate and
offering a unified perspective on experimentally observed diagonal features in various vdW systems.
We supported our analytical findings with mean-field calculations, 
determining layer potentials self-consistently, thereby reproducing experimental trends.

Our analysis focused on cases where the bottom-layer surface states dominate, corresponding to positive displacement fields that favor conduction electrons in the lower graphene layers. By symmetry, the same physics applies to the upper surface for negative displacement fields. Although we concentrated on twisted double-bilayer graphene, our framework can be extended to other multilayer graphene systems with local Bernal stacking, such as twisted bilayer-trilayer graphene, and rhombohedral multilayers ranging from thin films to bulk flakes. In particular, rhombohedral pentalayer graphene demonstrates a tendency towards single-gate tracking at small positive filling factors (e.g., Refs.~\cite{Lu2024,Lu2025,PhysRevX.15.011045}), occasionally coinciding in the $\nu-D$ plane with a region exhibiting robust correlations and quantum anomalous Hall states. The coexistence of the two effects potentially hints at a connection between strong correlations and single-gate tracking.

The self-consistently determined layer-potential framework may also be applied to scanning tunneling microscopy (STM) measurements versus gate bias, which corresponds to tracing a diagonal line in the displacement field and filling factor plane (effectively setting the density of carriers in the top gate to zero). The differing electrostatics between singly- and doubly-gated multilayer devices may help clarify the recent distinctions in symmetry-breaking regimes observed between transport and STM devices \cite{2024arXiv241111163L, 2024arXiv241114113L}. 

Lastly, other flat-band moir\'e systems, such as tMoTe$_2$ \cite{2024arXiv240609591P,2024arXiv240609687X} and tWSe$_2$ \cite{Xia2024, Guo2025}, also exhibit gate-tracking behavior in certain regions of their carrier density and displacement field parameter space. Our theory is likely able to capture similar effects in those systems, along with many other van der Waals multilayer structures.

\begin{acknowledgments}
We thank Felix von Oppen, Chunli Huang, Vo Tien Phong, and {\'E}tienne Lantagne-Hurtubise for useful discussions. We would also like to thank the HPC Service of ZEDAT, Freie Universit\"{a}t Berlin for computing time.
    KK was supported by Deutsche Forschungsgemeinschaft through CRC 183 (project C02), by a
joint ANR-DFG project (TWISTGRAPH, ANR-21-
CE47-0018), and by a grant from the
Simons Foundation (SFI-MPS-NFS-00006741-12, P.T.)
    in the Simons Collaboration on New Frontiers in Superconductivity.  D.W. was supported by an appointment to the Intelligence Community Postdoctoral Research Fellowship Program at University of Washington administered by Oak Ridge Institute for Science and Education through an interagency agreement between the US Department of Energy and the Office of the Director of National Intelligence. J.F. acknowledges support from the European Research Council (ERC) under the European Union Horizon 2020 research and innovation programme under grant agreement No. 951541; the Natural Sciences and Engineering Research Council of Canada; the Canadian Institute for Advanced Research; the Max Planck-UBC-UTokyo Centre for Quantum Materials and the Canada First Research Excellence Fund, Quantum Materials and Future Technologies Program.   M.Y. acknowledges support from NSF CAREER award no. DMR-2041972 and the Department of Energy, Basic Energy Science Programs under award DE-SC0023062. C.L. was supported by start-up funds from Florida State University and the National High Magnetic Field Laboratory. The National High Magnetic Field Laboratory is supported by the National Science Foundation through NSF/DMR-2128556 and the State of Florida.  \end{acknowledgments} 

    \IfFileExists{klibrary.bib}{\bibliography{klibrary.bib,additional.bib}}{\IfFileExists{../klibrary.bib}{\bibliography{../klibrary.bib}}{\bibliography{~/notes/3_moire/7_tmng/klibrary.bib}}}

    \clearpage

    \renewcommand{\appendixname}{Supplementary Information}

    \appendix

    \setcounter{equation}{0}
    \renewcommand{\thesection}{Supplementary Information \arabic{section}}    
    \renewcommand{\theequation}{S\arabic{section}.\arabic{equation}}
    \renewcommand{\thefigure}{S\arabic{figure}}
    \setcounter{figure}{0}
    \begin{widetext}
    \section{Details of the analytical model calculation}
    \subsection{Details on the single sector derivation}
    \label{sec:single_sector_derivation}
Upon varying the gate charges, the chemical potential changes (i) due to the compressibility $\invrholoc$ at constant $U_l$ and (ii) 
due to the change in layer potentials $\delta U_l$, which give additional shifts.
Assuming all the states at the Fermi level have the same distribution of charges $W_l$,  
the additional Fermi level shift due to the change in layer potentials 
is equal to the expectation of the change in the potentials $\delta U_l$ upon doping:
\begin{eqnarray}
-\frac{\partial \mu}{\partial \dens_b} &=&\invrholoc - \sum_l W_l \frac{\partial U_l}{\partial \dens_b}\\
-\frac{\partial \mu}{\partial \dens_t} &=&\invrholoc  - \sum_l W_l \frac{\partial U_l}{\partial \dens_t}, 
\end{eqnarray}
where $\frac{\partial U_l}{\partial \dens_b}$ and $\frac{\partial U_l}{\partial \dens_t}$ can be evaluated using
Eq.~\eqref{eq:potdifference}.
The change in the layer potentials upon changing the bottom gate charge by $\delta \dens_b$ is
\begin{equation}
\delta U_l =\frac{\partial U_l}{\partial \dens_b}\delta \dens_b= -\frac{e^2d}{\epsilon_\perp \epsilon_0} \left[(l-1)- \sum_{l'<l}(l-l')W_{l'}  \right] \delta \dens_b,
\end{equation}
while the change in the layer potentials upon changing the top gate charge by $\delta \dens_t$ is
\begin{equation}
\delta U_l =\frac{\partial U_l}{\partial \dens_t}\delta \dens_t= \frac{e^2d}{\epsilon_\perp \epsilon_0} \left[ \sum_{l'<l}(l-l')W_{l'}  \right] \delta \dens_t.
\end{equation}
Plugging this into Eqs.~\eqref{eq:somea} and \eqref{eq:someb} we obtain Eqs.~\eqref{eq:dmdnb},\eqref{eq:dmdnt}.
    \subsection{Multiple sectors}
    The analytical discussion of screening
    in the main text and above  assumed that all states at the Fermi level have the same distribution of charges $W_l$ across the layers. 
    In this Appendix, we generalize this analysis and derive the dependence of the chemical potential on the two gates for multiple sectors,
    and derive the result that for the \surface{} state evolution, the main text analysis is valid provided the parameters
    are understood to correspond to the delocalized (screening) states only, excluding the \surface{} pocket contribution.
    A special case of this analysis is having as many sectors as there are $k$ points in the Brillouin zone. For this reason we will consider different sectors labeled with an index $k$, where each sector has a layer distribution $W_l^{(k)}$, and density of states $\rholoc^{(k)}$.
    We now define a density of states matrix $\rhomat$ as
    \begin{equation}
    \rhomat_{kk} = \rholoc^{(k)},
    \end{equation}
    the matrix of layer spreads
    \begin{equation}
    \label{eq:labelspreadmatdef}
    \deltadmat_{ij} = \frac{e^2 d}{\epsilon_\perp\epsilon_0}\sum_{l'<l}W^{(i)}_l(l-l')W^{(j)}_{l'}.
    \end{equation}
    The diagonal components simplifies to $\deltadmat_{ii} =\frac{e^2 }{\epsilon_\perp\epsilon_0} \deltad^{(i)}$, where $\deltad^{(i)}$ is defined using the single-sector definition above, 
    given in Eq.~\eqref{eq:defdeltad}.
    The matrix of sector distances from the bottom layer is defined as
    \begin{equation}
    \dstarmat_{kj} = \frac{e^2 }{\epsilon_\perp\epsilon_0}d^{*,(k)}=
    \frac{e^2d }{\epsilon_\perp\epsilon_0}\sum_{l}(l-1)W^{(k)}_l, \end{equation}
    with $d^{*,(k)}$ being the distance from the bottommost layer of the sector $k$.
    Following a similar analysis as for the case of a single sector above, we obtain the following results
    for the gate projected compressibilities
    \begin{eqnarray}
    \label{eq:matrixdmudnb}
    -\left(\frac{\partial \mu}{\partial \dens_b}\right)^{-1} &=& \sum_{i,j}\left[ (\rhomat^{-1} - \deltadmat+\dstarmat)^{-1}\right]_{i,j}\\
        \label{eq:matrixdmudnt}
        -\left(\frac{\partial \mu}{\partial \dens_t}\right)^{-1} &=& \sum_{i,j}\left[ (\rhomat^{-1} - \deltadmat)^{-1}\right]_{i,j}.
        \end{eqnarray}

        In the weakly interacting limit $|\dstarmat| \ll |\rhomat^{-1}|$, $|\deltadmat| \ll |\rhomat^{-1}|$, we can obtain the inverses in Eqs.~\eqref{eq:matrixdmudnb},~\eqref{eq:matrixdmudnt}
        in a perturbative expansion. 
        For example, expanding the matrix inverse in Eq.~\eqref{eq:matrixdmudnt} obtains
        \begin{eqnarray}
        \label{eq:matrixdmudntexp}
        (\rhomat^{-1} - \deltadmat)^{-1} = 
        \rhomat + \rhomat\deltadmat   \rhomat + \mathcal{O}[\rhomat(\deltadmat \rhomat )^2],
        \end{eqnarray}
        such that the compressibility is in this limit equal to
        \begin{eqnarray}
        \label{eq:matrixdmudntapprox}
        -\frac{\partial \mu}{\partial \dens_t} &\approx& \frac{1}{\sum_k \rholoc^{(k)}} - \frac{1}{\sum_k \rholoc^{(k)}}\sum_{ij}  \rholoc_i (\deltadmat)_{i,j}\rholoc_j
        = 
        \invrholoc  - \frac{e^2 \deltad}{\epsilon_\perp\epsilon_0},
        \end{eqnarray}
        recovering the result of Eq.~\eqref{eq:dmdnt} upon identifying the total density of states $\rholoc = \sum_k \rholoc^{(k)}$ and 
        $\deltad$ as the average over the Fermi surface over all the sectors with sector $k$ weighted by $\frac{\rholoc^{(k)}}{\sum_k \rholoc^{(k)}}$.
        An analogous manipulation on Eq.~\eqref{eq:matrixdmudnb} 
        recovers Eq.~\eqref{eq:dmdnb} in the limit $|\dstarmat|\ll |\rhomat^{-1}|$ and $|\deltadmat| \ll |\rhomat^{-1}|$.

        We now apply this general framework to understand the evolution of the $\mu=0$ contour, at which the \surface{} state is at the Fermi level.
        This situation is naturally modeled using two sectors.
        One corresponding to the \surface{} states, fully polarized to the bottom layer, having $W_{l=1}^{(1)}=1$.
        The second sector corresponds to all the other states at the Fermi level, and has some layer-delocalized distribution $W_l^{(2)}$.
        In this limit, evaluating the compressibilities in Eq.~\eqref{eq:matrixdmudnb} and Eq.~\eqref{eq:matrixdmudnt} is
        straightforward, and we obtain
        \def\const{\mathfrak{C}}
        \begin{eqnarray}
        -\frac{\partial \mu}{\partial \dens_b} &=& \frac{1}{\const}\left[\frac{1}{\rholoc^{(2)}} - \frac{e^2 \deltad^{(2)}}{\epsilon_\perp\epsilon_0}+\frac{e^2 d^{*,(2)} }{\epsilon_\perp\epsilon_0}\right] \\
            -\frac{\partial \mu}{\partial \dens_t} &=&\frac{1}{\const}\left[\frac{1}{\rholoc^{(2)}}  - \frac{e^2 \deltad^{(2)}}{\epsilon_\perp\epsilon_0}\right],
        \end{eqnarray}
        where $\const = 1+ \frac{\rholoc^{(1)}}{\rholoc^{(2)}} +\frac{e^2\rholoc^{(2)} }{\epsilon_\perp\epsilon_0}(d^{*,(2)}-\deltad^{(2)})$.
        Remarkably, up to a rescaling by $\const$, these are the same results as for a single sector, but now all the quantities refer to sector $k=2$ only.
        This implies that $\alpha$ for the $\mu=0$ contour is 
        \begin{equation}
        \label{eq:suppldefalpha}
        \alpha \equiv \frac{\partial\mu}{\partial \dens_b}/  \frac{\partial\mu}{\partial \dens_t} = 1 + \frac{d^{*,(2)}}{\deltad^{(2)}}\left(\frac{\epsilon_\perp \epsilon_0}{e^2\deltad^{(2)}}\frac{1}{\rholoc^{(2)}}-1\right)^{-1}\,,
        \end{equation}
        which is the same as Eq.~\eqref{eq:defalpha}, but with all quantities defined with respect to sector $k=2$.
        Physically, this result arises since along the $\mu=0$ contour, the \surface{} states' occupation does not change -- their position with respect to the
        Fermi level is constant. Therefore, along the $\mu=0$ contour, all the charge enters the $k=2$ sector, meaning $\alpha$ should be defined
        with quantities referring to sector $k=2$ only.
        \subsection{Effect of imperfect layer polarization}
        \label{suppl:polar}
Supposing that the \surface{} states have a finite weight on the second layer $W^{lp}_2$, then to track the states,
we need to consider the evolution of $\mu-U_2W^{lp}_2$ as follows
\begin{eqnarray}
\frac{\partial (\mu-U_2W^{lp}_2)}{\partial \dens_t} &=&  \frac{\partial
\mu}{\partial \dens} \frac{\partial \dens}{\partial \dens_t} + \sum_{l} W_l
\frac{\partial U_l}{\partial \dens_t} - W^{lp}_2
\frac{\partial U_2}{\partial \dens_t}\label{eq:somea2}\\ 
\frac{\partial(\mu-U_2W^{lp}_2)}{\partial \dens_b} &=&  \frac{\partial \mu}{\partial \dens}
\frac{\partial \dens}{\partial \dens_b} + \sum_{l} W_l  \frac{\partial
U_l}{\partial \dens_b}
- W^{lp}_2\frac{\partial U_2}{\partial \dens_t}
\label{eq:someb2},
\end{eqnarray}
from which we obtain
\begin{eqnarray}
-\frac{\partial (\mu-U_2W^{lp}_2)}{\partial \dens_b} &=&\invrholoc - 
\frac{e^2 \deltad}{\tilde \epsilon_\perp\epsilon_0}+\frac{e^2 \dstar }{\tilde \epsilon_\perp\epsilon_0} 
+ (W_1-1) W^{lp}_2\frac{e^2 d }{\tilde \epsilon_\perp\epsilon_0}
\\
-\frac{\partial (\mu-U_2W^{lp}_2)}{\partial \dens_t} &=&\invrholoc  - \frac{e^2 \deltad}
{\tilde \epsilon_\perp\epsilon_0} 
+ W_1 W^{lp}_2\frac{e^2 d }{\tilde \epsilon_\perp\epsilon_0},
\end{eqnarray}
such that $\alpha$ becomes
\begin{equation}
\label{eq:defalphasupplie}
\alpha = -\frac{\delta \dens_b}{\delta \dens_t} = \frac{1 -\frac{e^2 \deltad}{\tilde\epsilon_\perp\epsilon_0}\rholoc+ W_1 W^{lp}_2\frac{e^2 d }{\tilde \epsilon_\perp\epsilon_0}\rholoc}
{ 1 +(\frac{\dstar}{\deltad}-1)\frac{e^2 \deltad}{\tilde\epsilon_\perp\epsilon_0}\rholoc + (W_1-1) W^{lp}_2\frac{e^2 d }{\tilde \epsilon_\perp\epsilon_0}\rholoc},
\end{equation}
so that $\alpha$ is increased upon adding the effect of $W^{lp}_2$ provided 
$W_1 + (1-W_1)\alpha>0$, which is satisfied provided 
$\alpha> \frac{W_1-1}{W_1} \approx -1 $.
Note that again in Eq.~\eqref{eq:defalphasupplie} all the quantities should be understood to refer to the screening states,
since we are interested in the $\mu-U_2W^{lp}_2=0$ contour.

        \section{Single particle models}
        For twisted $M+M'$ ($M+M'=N$) layer graphene we take the following kinetic Hamiltonian (without layer potentials):
\begin{equation}
\label{eq:spham1}
\hsp^{M+M'}(\mathbf k)=
\begin{pmatrix}
H_M(\mathbf k) & 0\\
0&H_{M'}(\mathbf k) 
\end{pmatrix}+ T_{M\to M+1} + T_{M+1\to M},
\end{equation}
where
$H_M$ is the Hamiltonian of a Bernal stacked $M$-layer \cite{falkoMcCannLandauLevelDegeneracyQuantum2006,peresNilssonElectronicPropertiesBilayer2008,koshinoMcCannElectronicPropertiesBilayer2013}:
\begin{equation}
\label{eq:spham2}
H_{M}=
\begin{pmatrix}
v_F \mathbf k \cdot \mathbf \sigma & t^\dagger(\mathbf k)&0&\cdots\\
t(\mathbf k) &v_F \mathbf k \cdot \mathbf \sigma & t(\mathbf k)&\cdots\\
0&t^\dagger(\mathbf k) &v_F \mathbf k \cdot \mathbf \sigma  &\cdots\\
\vdots & \vdots &\vdots &\ddots
\end{pmatrix},
\end{equation}
with $v_F$ the Dirac velocity and $t(\mathbf k) = 
\begin{pmatrix}
-v_4 k & t_1 \\
-v_3 \overline k & -v_4 k
\end{pmatrix}$, with $v_3,v_4$ the non-local tunneling strengths and $t_1$ the strength of AB hopping when A is on top of B.
We use the following parameters
$v_F = \SI{542.1}{meV\cdot nm}$, $v_3 =v_4 = \SI{34}{meV\cdot nm}$, 
$t_1 =\SI{355.16}{meV}$ \cite{yuHerzog-ArbeitmanMoireFractionalChern2023}.
The moir\'e tunneling term is $\left.T_{M+1\to M}\right|_{l',l}= T(\mathbf r)\delta_{l',M+1}\delta_{l,M} = T^\dagger_{M\to M+1},$
where $T(\mathbf r) = \sum_{j=0}^2 T_j e^{i\mathbf q_j\cdot \mathbf r}$ is the interlayer hopping with $T_j = w_{AA}\sigma_0 + w_{AB} \left[\sigma_x \cos(2\pi j/3)+
\sigma_y \sin(2\pi j/3)\right] $ and $\mathbf q_j = (O_3)^j (\mathbf{K}_{M+1}-\mathbf{K}_{M}) = 2|K| \sin\left(\theta/2\right) (O_3)^j  \left[0 ,-1\right] $ with $\mathbf{K}_i$ the K-point positions in layer $i$ 
and $O_3$ the matrix of a counterclockwise $120^\circ$ rotation.
We take $w_{AB} =\SI{110}{meV}$ \cite{macdonaldBistritzerMoireBandsTwisted2011} and
$w_{AA} =0.7w_{AB}= 0.7 \cdot \SI{110}{meV}$, with the factor $0.7$ accounting for lattice corrugation.
For the single particle analysis, we add a layer potential term $H_{\perp}$ with potentials given by $U_l = \Delta U (l-1)$.
For twisted double bilayer graphene, we have specifically:

\begin{equation}
\label{eq:spham3}
\hsp^{TDBG}=
\begin{pmatrix}
v_F \mathbf k \cdot \mathbf \sigma & t^\dagger(\mathbf k)&0&0\\
t(\mathbf k) &v_F \mathbf k \cdot \mathbf \sigma & T(\mathbf r)&0\\
0&T^\dagger(\mathbf r) &v_F \mathbf k \cdot \mathbf \sigma  &t^\dagger(\mathbf k)\\
0&0 & t(\mathbf k)&v_F \mathbf k \cdot \mathbf \sigma  
\end{pmatrix}.
\end{equation}

\section{Internal Bernal stacking interfaces}

Here we discuss the case in which an internal interface has Bernal stacking.
Suppose that layer $l$ is sandwiched between surrounding layers in an ABA stacking. In the basis $\Psi_{\mathbf k} = (\ldots,c_{\mathbf k,Al-1},c_{\mathbf k,Bl-1},c_{\mathbf k,Al}, c_{\mathbf k,Bl}, c_{\mathbf k,Al+1},c_{\mathbf k,Bl+1}, \dots)$, 
the Hamiltonian, focusing on the $Al$ state then reads
\begin{equation}
\hsp + H_\perp=
\sum_{\mathbf k}
\Psi^\dagger_{\mathbf k}
    \begin{pmatrix}
        &&\cdots &&\cdots \\
        &&-v_4 k &&\cdots \\
        & \ddots& -v_3 \overline{k}& \multicolumn{5}{c}{\multirow{5}{*}{$\displaystyle \cdots $}}\\
        -v_4 \overline{k}&-v_3 k &U_l &  v_F \overline{k}&-v_4 \overline{k}&-v_3 k &0&\cdots \\
        && v_F k&    & \cdots \\
        \vdots&& -v_4 k&\\
        &&-v_3 \overline{k}& \\
        & \\
    \end{pmatrix}
\Psi_{\mathbf k},
\end{equation}
where the $Al$ state is seen to be an exact eigenstate of energy $U_l$. 
Choosing a gauge where $U_l=0$, this state
is at the Fermi level when $\mu=0$. 
In the M+N layer devices considered \cite{yankowitzWatersTopologicalFlatBands2024}, the relevant state
for the low energy physics appears to be the state at layer $M-1$ when the conduction band is polarized towards the $M$ layer Bernal component, and
our analysis in the main text naturally carries over.

\section{Details on the self-consistent numerics}

For the mean-field calculation, we choose at each momentum $\mathbf k$ on a Brillouin zone grid a basis of $N_{\mathbf G}$ plane waves
per layer and sublattice, and use the same basis set for each value of displacement field and filling.
In this method, the remote bands that were not included in the calculation can be captured
by using a larger $\epsilon_\perp$, which is displacement field independent. On the other hand, for a basis that depends on displacement field, obtained
typically by diagonalizing the single particle Hamiltonian and keeping $N_{active}$ central bands,
$\epsilon_\perp$ would have to depend on displacement field, and taking a fixed value for it would incur a displacement-field dependent bias.

To determine $U_l$ in the out-of-plane term $H_\perp$, 
we need to use the net charge density relative to charge neutrality $\dens_l$. We obtain obtain it by 
subtracting from the electrons present in our mean-field state in each layer their number at graphene charge neutrality. For the finite plane wave set, this  gives $N_f \cdot N_{\mathbf G}$ per momentum $\mathbf k$ in the Brillouin zone.

For the $\mathbf q\neq 0 $ term of the interaction, we use
\begin{equation}
H_{\text{int},\mathbf{q} \neq 0} = \frac{1}{2A} \sum_{\mathbf{q}\neq 0}\sum_{i,j} V_{ij}(\mathbf q) :\rho_{\mathbf q,i} \rho_{-\mathbf q,j}:,
\label{eq:interactinghamiltonian}
\end{equation}
where $A$ is the system area, $::$ denotes normal ordering, and $\rho_{-\mathbf q,j}$ is the charge density in layer $j$ at momentum $\mathbf q$,
and where $V_{ij}(\mathbf q)$ is the double-gate screened Coulomb interaction accounting for the layer index \cite{lewandowskiKolarElectrostaticFate$N$layer2023},
given as 
\begin{align} 
\label{eq:doublegatecoulomb}
 V_{i,j}(\mathbf q)  && =
&&\frac{1}{2\epsilon_\parallel \epsilon_0}\frac{1}{q}\cdot \left(\frac{e^{-q (z_i+z_{j})} 
\left(-e^{2 q (d_s+z_i+z_{j})}-e^{2 d_s q}+e^{2 q z_i}+e^{2 q z_{j}}\right)}{e^{4 d_s q}-1}+e^{-q |z_i-z_{j}|} \right),
\end{align} 
where $\epsilon_\parallel$ is the in-plane dielectric constant, $z_i$ is the position of layer $i$, $d_s$ is the vertical distance from the center of the sample to the top/bottom gate,
which we set to be $d_s= \SI{40}{nm}$ throughout. Furthermore, we set $\epsilon_\parallel =6$.

For the symmetry unbroken calculations, we treat the $\mathbf{q} \neq 0$ term in the Hartree approximation.
In particular, we solve $\hat{H} = \hsp + H_\perp+ \H{Hartree}$, where the Hartree term reads 
\begin{equation}
 \label{eq:hmfhartree}
\H{Hartree} =\frac{1}{A}\sum_{i,j}\sum_{\mathbf G}\rho_{\mathbf G,i} V_{i,j}(\mathbf G) \left\langle 
\rho_{-\mathbf G,j}\right \rangle,\end{equation}
where $\mathbf G$ are moir\'e reciprocal lattice vectors, 
and where $\left\langle \rho_{-\mathbf G,j}\right \rangle$ denotes the expectation value of the momentum $-\mathbf G$ density in the mean-field state.
We only simulate for one flavor and to obtain results for $N_f=4$ spin-valley flavors, we multiply all the interacting terms by $N_f$, modeling unbroken flavor symmetry.

To study flavor symmetry breaking, we add the term of Eq.~\eqref{eq:defhfock}, meaning we solve
$\hat{H} = \hsp + H_\perp+ \H{Hartree}+\H{Fock}$ self-consistently, considering independent filling of the different spin flavors.
To define the region in which cascade happens, 
\def\cascmeasure{\mathcal{F}}
we consider for $\nu<2$ a measure $\cascmeasure = \left|\frac{\nu_{K\downarrow}+\nu_{K'\downarrow}-\nu_{K\uparrow}-\nu_{K'\uparrow}}{\nu}\right|$, 
which ranges from $\cascmeasure=0$ for no symmetry breaking to $\cascmeasure=1$ for a full cascade.
For $\nu>2$, to quantify symmetry breaking, we consider the filling of holes, 
so in the expression for $\cascmeasure$, one needs to take $\nu_f \to 1-\nu_f $.
The region in which cascade occurs is then defined by $\cascmeasure\geq 0.5$. Note that at this Hartree level with our unbiased basis set,
the choice of a subtraction scheme is unimportant \cite{bernevigKwanMoireFractionalChern2023}.
We use the optimal damping algorithm \cite{lebrisCancesCanWeOutperform2000} to obtain convergence.

\section{Extended data}

\begin{figure}[t]
    \includegraphics[width=0.9\columnwidth]{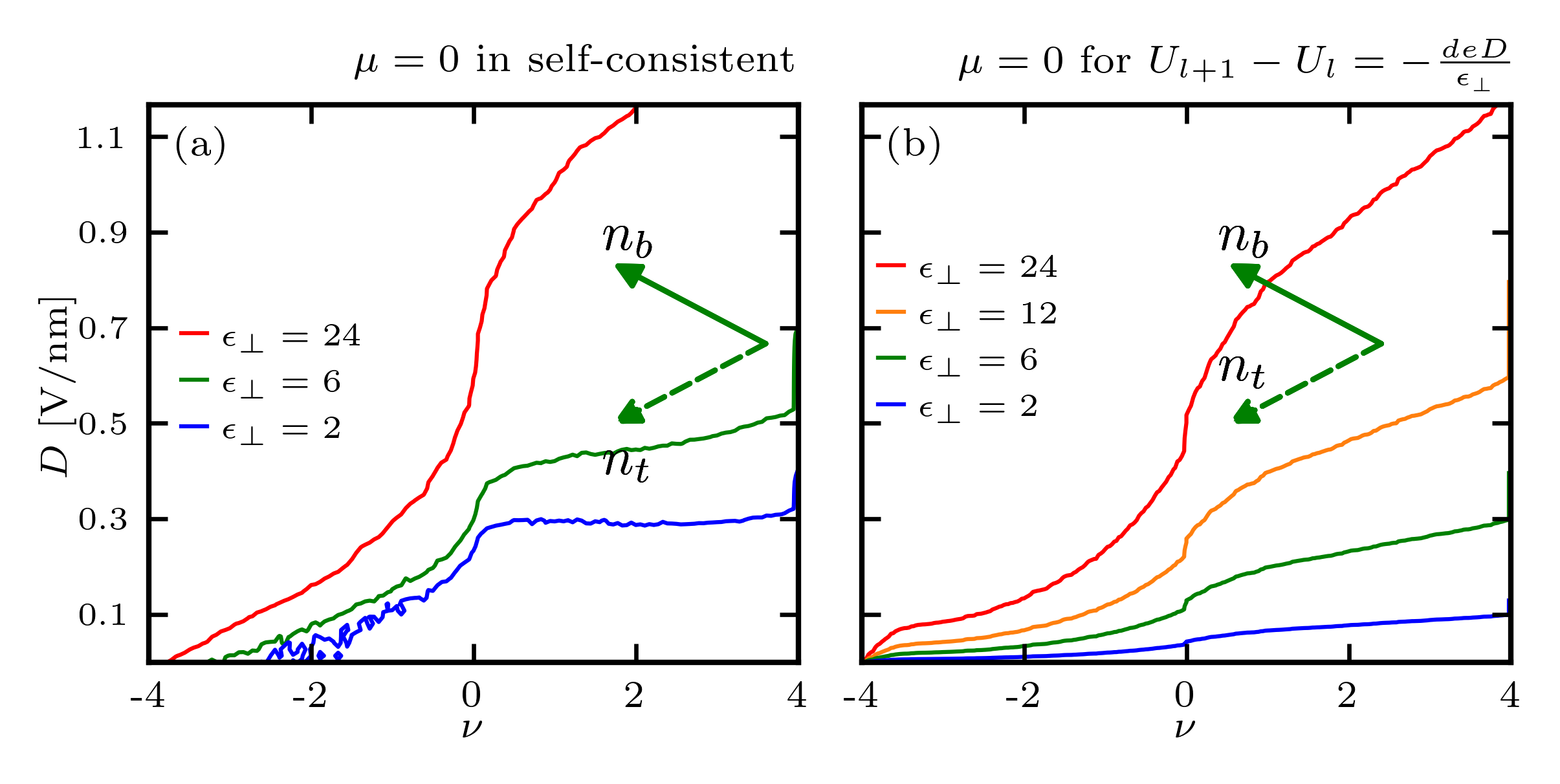}
    \caption{(a) $\epsilon_\perp$ dependence of $\mu=0$ contour for $\epsilon_\parallel=6$ within the self-consistent Hartree calculation. 
    (b) $\epsilon_\perp$ dependence of the $\mu=0$ contour in the SP calculation assuming $\Delta U =  -deD/\epsilon_\perp$.}
    \label{fig:figsuppla}
\end{figure}
Fig.~\ref{fig:figsuppla}a shows the $\epsilon_\perp$ dependence of the $\mu=0$ contour in the $\nu,D$ plane. 
Importantly, the $\nu<0$ gate tracking is robust for $\epsilon_\perp =2,6,24$.
The importance of self-consistent layer potentials can be appreciated by naively attempting to convert $\Delta U$ to $D$ using $\Delta U =  -deD/\epsilon_\perp$
in the single particle calculation of Sec. IIA of the main text, which we show in Fig.~\ref{fig:figsuppla}b.

We show the additional cascade data in Fig.~\ref{fig:figsupplb}.
In Fig.~\ref{fig:figsupplb}a, we show the cascade for $\intstrengthfock=\SI{6.7}{meV}$ at $\epsilon_\perp=6$, which looks similar to Fig.~4a in the main text. 
In particular, the cascade boundaries closely correspond to the \surface{} pocket depletion.
On the other hand, at a stronger Fock, $\intstrengthfock=\SI{8.3}{meV}$, shown in Fig.~\ref{fig:figsupplb}b,
the cascade boundaries are not directly related to the \surface{} pocket, 
as can be also seen by observing the density of states in Fig.~\ref{fig:figsupplb}c. This is in line with our understanding of
the single gate tracking transitions as arising from an intermediate coupling effect, in which the density of states 
due to the \surface{} pocket is important.
In Fig.~\ref{fig:figsupplb}d, we show the cascade for $\intstrengthfock=\SI{6.7}{meV}$ at $\epsilon_\perp=2$.
While the cascade boundaries deviate from the \surface{} state occupation $N_{pol}$ boundaries, this deviation arises because the depletion
of the \surface{} pocket happens at finite $\mathbf k$, and does not exactly coincide with the \surface{} depletion. 
These finite $\mathbf k$ states have a small weight on the second layer, and as shown in~\ref{suppl:polar},
this is expected to increase the value of $\alpha$, in this case making it less negative.

At stronger interactions (Fig.~\ref{fig:figsupplb}e,f), the \surface{} pocket again becomes less relevant, as the details
of the band structure become unimportant and symmetry breaking is abundant.

\begin{figure}[t]
    \includegraphics[width=\columnwidth]{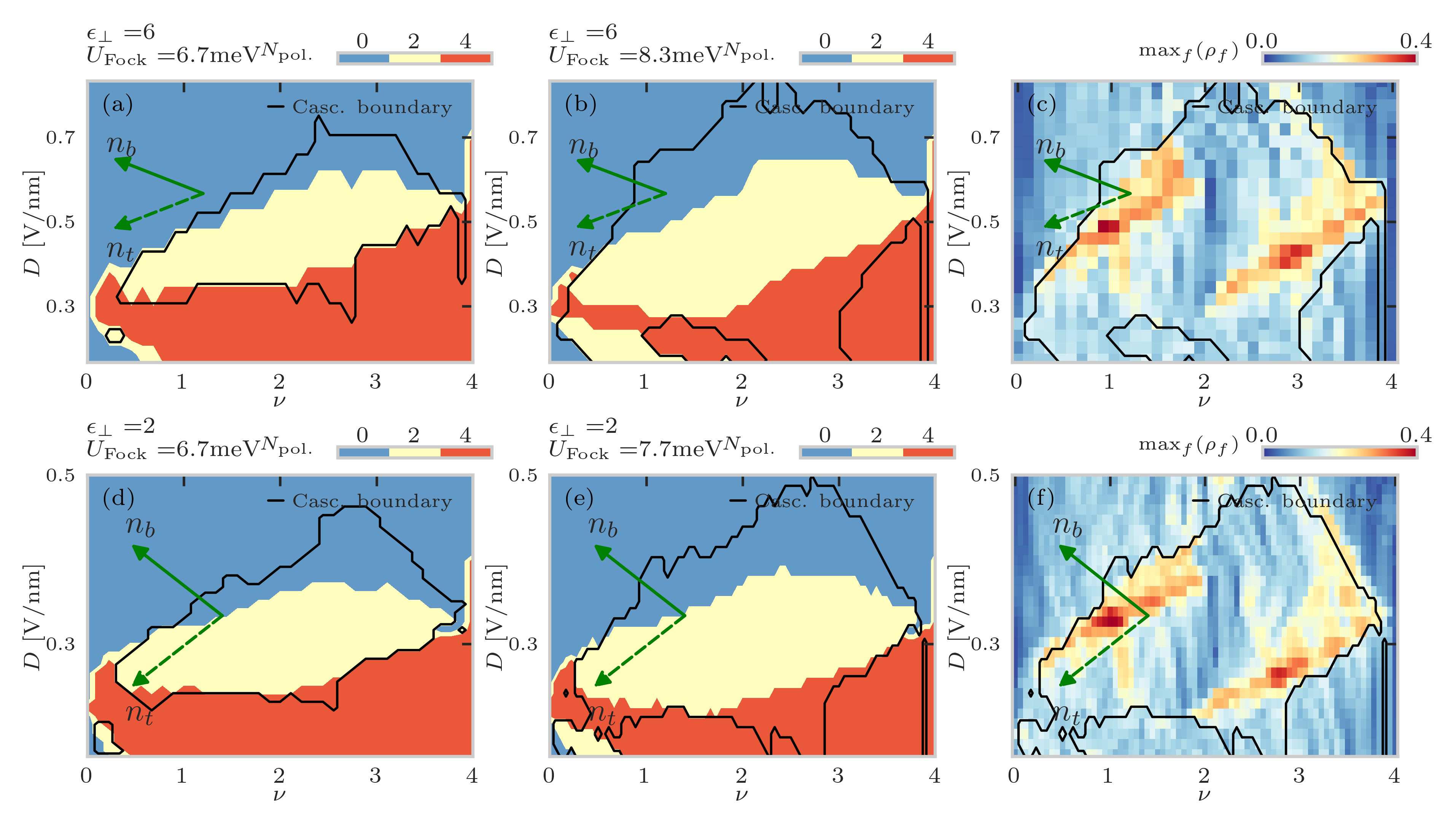}
    \caption{Cascade plots for other values of $\intstrengthfock$ and $\epsilon_\perp=2$ than in the main text. We work at $\epsilon_\parallel=6$ throughout for the in-plane Hartree interaction.}
    \label{fig:figsupplb}
\end{figure}
\end{widetext}

\end{document}